\newcommand{\ud}{\rm d}
\newcommand{\un}{~\mathrm}
\newcommand{\ie}{{\em i.e. }}
\documentclass[twocolumn
,floatfix
,showpacs
,preprintnumbers
,amsmath
,amssymb
,superscriptaddress]{revtex4}

\usepackage[T1]{fontenc}
\usepackage{graphicx}
\usepackage{dcolumn}
\usepackage{bm}

\begin{document}

\title{Glass breaks like metals, but at the nanometer scale}
\author{F. Célarié}
\affiliation{Laboratoire des Verres - UMR CNRS-UM2 5587, Université Montpellier 2, C.C. 69 - Place Bataillon, F-34095 Montpellier Cedex 5 - France}
\author{S. Prades}
\affiliation{Service de Physique et Chimie des Surfaces et Interfaces, DSM/DRECAM/SPCSI, CEA Saclay, F-91191 Gif sur Yvette - France}
\author{D. Bonamy}
\affiliation{Laboratoire des Verres - UMR CNRS-UM2 5587, Université Montpellier 2, C.C. 69 - Place Bataillon, F-34095 Montpellier Cedex 5 - France}
\affiliation{Service de Physique et Chimie des Surfaces et Interfaces, DSM/DRECAM/SPCSI, CEA Saclay, F-91191 Gif sur Yvette - France}
\author{L. Ferrero}
\affiliation{Laboratoire des Verres - UMR CNRS-UM2 5587, Université Montpellier 2, C.C. 69 - Place Bataillon, F-34095 Montpellier Cedex 5 - France}
\author{E. Bouchaud}
\affiliation{Service de Physique et Chimie des Surfaces et Interfaces, DSM/DRECAM/SPCSI, CEA Saclay, F-91191 Gif sur Yvette - France}
\author{C. Guillot}
\affiliation{Service de Physique et Chimie des Surfaces et Interfaces, DSM/DRECAM/SPCSI, CEA Saclay, F-91191 Gif sur Yvette - France}
\author{C. Marlière}
\affiliation{Laboratoire des Verres - UMR CNRS-UM2 5587, Université Montpellier 2, C.C. 69 - Place Bataillon, F-34095 Montpellier Cedex 5 - France}

\begin{abstract}

We report in situ Atomic Force Microscopy experiments which reveal the presence of nanoscale damage cavities ahead of a stress-corrosion crack tip in glass. Their presence might explain the departure from linear elasticity observed in the vicinity of a crack tip in glass. Such a ductile fracture mechanism, widely observed in the case of metallic materials at the micrometer scale, might be also at the origin of the striking similarity of the morphologies of fracture surfaces of glass and metallic alloys at different length scales.

\end{abstract}


\date{\today}
\maketitle


Glasses are the most common example of "brittle" materials which break abruptly, without first deforming in an irreversible way as metals do. In metallic alloys, cracks usually progress through the coalescence of damage cavities, which nucleate within microstrutural defects (second phase precipitates, grain boundaries...) or at the interface between the matrix and the heterogeneities. This ``ductile" fracture mode, widely observed for a large variety of metallic alloys~\cite{Pineau95}, leads to very rough fracture surfaces which have been extensively studied over the last eighteen years~\cite{Mandelbrot84,Bouchaud97}. For slow crack propagation, the fracture surfaces of glass may appear very flat (in the so-called "mirror" zone) if examined with an optical microscope. However, when analysed at the nanometer scale with an Atomic Force Microscope (AFM), they reveal a roughness which is strikingly similar to the one exhibited by metallic fracture surfaces~\cite{Bouchaud97,Daguier95}. The only difference actually resides in the length scales involved, which are several orders of magnitude smaller in the case of glass. Could it be because, despite the conventional belief, the fracture mechanisms of these two categories of materials are similar, although taking place at different length scales? We report here the first experimental evidence of a ductile fracture mode in a vitreous material at a temperature much lower than the glass transition temperature Tg. Such scenario was predicted by Molecular Dynamics simulations~\cite{Nakano99,vanBrutzel97}, but it had never been observed experimentally up to now. The experiments reported here clearly show that slow fracture in glass progresses through the nucleation, growth and coalescence of damage cavities at the nanometer scale. These cavities are shown to be correlated to the non-linear elastic zone observed in the vicinity of the crack tip~\cite{Guilloteau96}. Possible origins of the cavities' nucleation are conjectured and the consequences of such ductile fracture mode in glass are discussed.

{\em Experimental setup.} The experimental setup is illustrated in Fig.~\ref{fig1}. All the experiments are performed at a constant temperature of $22.0\pm 0.5~^\circ\mathrm{C}$ and in a leak-proof chamber under an atmosphere composed of pure nitrogen and water vapor at a relative humidity of $42\pm 1\%$ after preliminary out-gassing.  Fracture is performed on DCDC~\cite{He95} (Double Cleavage Drilled Compression) parallelepipedic ($4\times4\times40\un{mm}^3$) samples of aluminosilicate glass. A thermal treatment ($660^\circ\mathrm{C}$) was performed before each fracture experiment in order to remove residual stresses~\cite{Marliere02}. The $4\times40\un{mm}^2$ surfaces are optically polished (RMS roughness is $0.25\un{nm}$ for a $10\times10~\mu\mathrm{m}^2$ scan size). In the centre of two parallel $4\times40\un{mm}^2$ surfaces and perpendicularly to them a cylindrical hole (radius $a=0.5\un{mm}$) is drilled. Its axis defines the $z$-direction. The $x$-axis (and $y$-axis) are parallel to the $40\un{mm}$ (and $4\un{mm}$) long side of the $4\times40\un{mm}^2$ surface. A compressive load is applied perpendicularly to the $4\times4\un{mm}^2$ surfaces. The external stress $\sigma$ is gradually increased by the slow constant displacement ($0.02\un{mm/min}$) of the jaws of the compression machine. Once the two cracks (symmetric to the hole axis) are initiated, the jaws'displacement is stopped. The crack front then propagates along the $x$-axis in the symmetry plane of the sample parallel to the ($x$,$z$) plane. In this geometry, the stress intensity factor $K_I$ is given~\cite{He95} by: $K_I=\sigma\sqrt{a}/(0.375c/a+2)$, where $c$ is the length of the crack (Fig.~\ref{fig1}a).

At the very first moments, the crack propagates very quickly. In this regime, the crack velocity $v$ is independent of the chemical composition of the surrounding environment~\cite{Wiederhorn67}. As the crack length $c$ increases, $K_I$ decreases, and $v$ decreases quickly. Under vacuum, the crack stops for $K_I$ smaller than a critical value $K_{Ic}$ referred to as the toughness of the material. But in a humid atmosphere, the corrosive action of water on glass allows for slow crack propagation at much lower values of the stress~\cite{Wiederhorn67}. The crack motion within the external ($x$,$y$) sample surface is then slow enough to be monitored by our experimental system combining optical microscopy and AFM (Fig.~\ref{fig1}b). Optical image processing gives the position of the crack tip and consequently the ``instantaneous" velocity for $v$ ranging from $10^{-6}$ to $10^{-9}\un{m.s}^{-1}$. By AFM measurements -~performed in a high amplitude resonant mode ("tapping" mode)~-, one probes the crack tip neighbourhood at magnifications ranging from $75\times75\un{nm}^2$ to $5\times5~\mu\mathrm{m}^2$~\cite{Marliere01} and the crack tip motion at velocities ranging from $10^{-9}$ to $10^{-12}\un{m.s}^{-1}$. The data presented below are obtained for  $K_I=0.43\mathrm{MPa.m}^{1/2}$ and $v=3.10^{-11}\un{m/s}$.

{\em Evidence of nano-scale damage cavities.} Typical topographical frames in the neighbourhood of the crack tip are presented in Fig.~\ref{fig2}. They clearly reveal cavities of typically $20\un{nm}$ in length and $5\un{nm}$ in width ahead of the crack tip (Fig.~\ref{fig2}a). These cavities grow with time (Fig.~\ref{fig2}b) until they coalesce (Fig.~\ref{fig2}c).

To ensure that the spots observed ahead of the crack tip are actually damage cavities which grow further and coalesce with the main crack leading to failure, we use the Fracture Surface Topography Analysis (FRASTA) technique~\cite{Kobayashi87} first introduced to study damage in metallic alloys. It consists in analyzing the mismatch between the two fracture surfaces. To understand how the FRASTA technique can provide physical details on the fracture mechanism, let us now consider how a {\em ductile} material breaks: In such a medium, the load application {\em first} induces a local plastic flow before generating any local failure at the level of the stress concentrators. {\em Then}, when local failure actually occurs,  the stress applied on the newly formed void surfaces vanish, the applied load is redistributed to nearby unbroken material, and plastic deformation is not undergone anymore at the level of these void surfaces. Consequently, in such a ductile scenario, each cavity initiation is accompanied by local irreversible plastic deformations {\em printed} in relief on the developing fracture surfaces (the crack lines when the method is applied in two dimensions as in the present case) that should remain visible after the cavities have coalesced and the crack has crossed.

The method consists in placing the upper fracture surface (crack line in 2D) under the lower one until no void is left. Then, the two surfaces (or lines) are pulled away from each other along the direction perpendicular to the fracture plane (or direction). This is what happens during the fracture process when the external strength is applied at a constant displacement rate. At the small length scales considered here, this assumption can be made. Cavities therefore appear in the {\em chronological} order.

The crack lines are first determined by binarising the image of the sample after fracture (Fig.~\ref{fig3}a), and the unbroken material is reconstituted virtually by placing numerically the lower line over the upper one (Fig.~\ref{fig3}b). The lower crack line is gradually translated in the direction of decreasing $y$ (Fig.~\ref{fig3}b), and the cavities appear and grow in the chronological order. The structure obtained for a given displacement, \ie at a given time, is superimposed on images recorded prior to failure and shown to correspond actually to cavities observed at this given time (Fig.~\ref{fig3}c). This indicates that the spots observed prior to failure are indeed depressions which are marked in relief on the final crack, and hence, are actual damage cavities.

{\em Displacement field.} A consequence of this "nano-scale ductility" can be seen in the displacement field around the crack tip (see also references~\cite{Guilloteau96} for related discussion): For a slit-like plane crack in an ideal Hookean continuum solid, the stress components $\sigma_{ij}$ at a given point M whose cylindrical polar coordinates are $(r,\theta,z)$ in the vicinity of the crack tip take the form $\sigma_{ij}(r,\theta,z)=f_{ij}(\theta)K_I\sqrt{2\pi r}$ where the functions $f_{ij}$ are completly determined from linear elastic theory~\cite{Irwin57}. Moreover, the $z$-displacement $u_z$ of the specimen surface (flat when no load is applied) is given by $u_z(r,\theta)=\int_{-h}^0\epsilon_{zz}(r,\theta,z)\ud z$ where $h=4\un{mm}$ is the thickness of the specimen and $\epsilon_{zz}$ is the direct strain in the $z$-direction. If the linear elastic stress-strain relation held on the nanometer scale, one gets $\epsilon_{zz}=(\sigma_{zz}-\nu(\sigma_{xx}+\sigma_{yy}))/E$ (where $E$ and $\nu$ are respectively the Young Modulus and the Poisson coefficient of the material) and consequently $u_z(r,\theta)=G(\theta) h \nu K_I/(E\sqrt{2\pi r})$, where $G(\theta)$ is a function equal to $G(\theta)=2\cos(\theta/2)$ for mode I fractures~\cite{Lawn93}.

Measurements of $u_z$ profiles have been performed on $1\times1~\mu\mathrm{m}^2$ AFM topographical frames (Fig.~\ref{fig4}a) along the direction of crack propagation (Fig.~\ref{fig4}b) and perpendicularly to it (Fig.~\ref{fig4}c). For both profiles, $u_z$ departs from the linear elastic $r^{-1/2}$ scaling for $r$ smaller than a threshold $r_c$ highly dependent on $\theta$: for $\theta=0^\circ$, $r_c=100\un{nm}$ while for $\theta=90^\circ$, $r_c=20\un{nm}$. These short-range departures from the linear elastic behaviour may be related to the presence of cavities although many other phenomena could be responsible for this discrepancy. However, the fact that the order of magnitude of the ratio $r_c(\theta=0^\circ)/r_c(\theta=90^\circ)$ -~much higher than the ratio of the cosine terms in the linear elastic expression of $u_z$~- is close to the aspect ratio of the observed damage cavities, strongly suggests a correlation between damage and non linear elasticity.

{\em Discussion.} Similar fracture experiments performed in amorphous Silica specimens reveal similar damage cavities. This suggests that their existence does not depend on the precise chemical composition of the studied glass. The cavities' nucleation should be found more likely in the amorphous structure, which contains inherent atomic density fluctuations at the nanometer scale. Such atomic density fluctuations have been evidenced by van Brutzel, Rountree, {\em et al.}~\cite{vanBrutzel97} in the structure of simulated amorphous silica by Molecular Dynamics: The Si and O atoms are shown to form Silica tetrahedra connected together to build rings of different sizes ranging from 3 to 9 tetrahedra. At larger length scales, ranging from $1.5\un{nm}$ to $6\un{nm}$, the density of these rings is found to fluctuate with high density areas surrounded by low density areas. Moreover, the Molecular Dynamics of van Brutzel~\cite{vanBrutzel97} show that, at this length scale, crack propagates by growth and coalescence of small cavities which appear in areas with low density of rings, ahead of the crack tip. They behave as stress concentrators and grow under the stress imposed by the presence of the main crack to give birth to the cavities actually observed in the AFM frames.
 
Another scenario was also proposed by Falk~\cite{Falk99,Damien02}: A small change in the interactomic potential was shown to be sufficient to generate a ductile to brittle transition in a 2D polydisperse packing of beads interacting through Lennard-Jones potential~\cite{Falk99}. This sensitivity was related to the existence of non-affine rearrangements zones (groups of atoms whose motion cannot be approximated by a linear local strain tensor) whose density depends strongly on the interatomic potential range. Unfortunalty, the AFM resolution is not presently sufficient to directly evidence these non-affine molecular rearrangements.

Moreover, two roughness regimes have been actually observed on post-mortem glass fracture surfaces~\cite{Bouchaud97,Daguier95}. The low length scales regime (from $1\un{nm}$ to a crossover length $\xi_c$ ranging from $10$ to $30\un{nm}$ depending on the average crack velocity) was interpreted as the consequence of the extension of an isolated damage cavity; the regime at larger length scales (from $\xi_c$ to $100\un{nm}$) is thought to be linked to the structure composed of the various correlated damage cavities~\cite{Bouchaud02}. The order of magnitude of $\xi_c$ is in good agreement with our present observations, where cavities at coalescence are a few tens of nanometers wide.

Here, let us note that AFM observations are performed on the sample surface, where the mechanical state is different from that of the bulk. It must be remembered that the stress components at the surface are contained within the free surface plane, while the deformations are fully three-dimensional. On the contrary, the bulk is in a condition of plane strain and fully three-dimensional stresses. Hence, the observed sizes and growth rates of cavities at the surface may well differ from those in the bulk. New experiments using the FRASTA method in three dimensions applied to the post-mortem study of the fracture surfaces are currently being performed, in order to have access to the three dimensional structure of bulk damage and its evolution. Through this new set of experiments, one should be able to correlate also the damage structure to the fracture surface morphology.

Moreover, the structure of damage, which influences macroscopic mechanical properties such as fracture toughness and lifetime, should then be linked to the glass composition and nanostructure~\cite{vanBrutzel97,Falk99,Benoit00}. Complementary analyses addressing the question of the chemical bonds on the fracture surface will also be performed.

Finally, the fact that glass at temperatures far below the glass transition temperature Tg, considered up to now as the archetype of pure brittle material, joins the class of damageable materials should have important consequences for its mechanical properties. In applications, the design of structures using glass might be modified to take this behaviour into account, especially for slow crack propagation processes.

The similarity between the damage modes of materials as different as glass and metallic alloys is an important clue to understand the origin of puzzling universal behaviors, hence sheding new light into the basic physical mechanisms of fracture.

\begin{acknowledgments}
We thank Ian Campbell and Nick Barrett for a critical reading of the manuscript. We also acknowledge Bernard Delettre for the making of the glass samples and Jean-Marie Felio and Emmanuel Arnould for technical support. We are indebted to Rajiv Kalia and Laurent Van Brutzel for enlightning discussions, and to Jean-Marc Cavedon and Georges Lozes for their constant support.

\end{acknowledgments}

\begin{figure*}
\centering
\caption{ experimental setup. (a): Sketch of the DCDC geometry (b): Picture of the experiment.} 
\label{fig1}
\end{figure*}

\begin{figure*}
\centering
\caption{Sequence of successive topographic AFM frames showing the crack propagation at the surface of the specimen. The scan-size is $75\times75\un{nm}^2$ and the heights range over $2\un{nm}$. The recording time for one frame is around $3\un{mn}$ and two successive frames are separated by $30\un{mn}$. The crack front propagates from the left to the right ($x$-direction) with a mean velocity $v$ of around $10^{-11}\un{m/s}$. (a): evidence of nanometric damage cavities before the fracture advance. (b): growth of the cavities. (c): the crack is advancing via the coalescence of all the cavities} 
\label{fig2}
\end{figure*}

\begin{figure*}
\centering
\caption{Fracture Surface Topographical Analysis (FRASTA). (a) Frame~\ref{fig3}c (broken sample) is binarised and the contours of the crack are determined. (b) The lower line is first numerically raised over the upper one and then gradually displaced in the direction of decreasing $y$, as schematized by the arrow. Cavities are coloured in red. (c): Result of the method: superimposition of the obtained cavities on the image~\ref{fig3}b recorded prior to complete failure.} 
\label{fig3}
\end{figure*}

\begin{figure*}
\centering
\caption{Measurements of the surface deformations and comparison with the predictions for an ideal Hookean material. The crack propagates from left to right ($x$ positive). (a): typical AFM topographical frame of the vicinity of the crack tip. The scan-size is $1\times1~\mu\mathrm{m}^2$ and the heights range over $3\un{nm}$. The white vertical (respectively horizontal) dotted line sets the $x$-coordinate $x_f$ (respectively the $y$-coordinate $y_f$) of the crack tip. (b) (respectively (c)) top: Plot of the $z$ profile along (respectively perpendicularly to) the direction of crack propagation. The open circles correspond to experimental data while the full line corresponds to the prediction $z=z_0-Ar^{-1/2}$ (where $z_0$ and $A$ are fit parameters) given for an ideal Hookean solid. Bottom: Log-log plot of the displacement $u_z=z_0-z$ versus the distance $r=x-x_f$ (respectively $r=y-y_f$) from the crack tip. For $r\leq r_c$, the $\delta$-profile departs from the predictions given by the linear elastic theory.} 
\label{fig4}
\end{figure*}


\begin{thebibliography}{99}

\bibitem{Pineau95} A. Pineau, D. François and A.  Zaoui, {\em Comportement mécanique des matériaux}, (Hermès, Paris, 1995);
E. Bouchaud and F. Paun, Computing in Sci. and Eng., pp. 32-38, (September/ October 1999).

\bibitem{Mandelbrot84} B.B. Mandelbrot, D.E. Passoja and A.J. Paullay, Nature {\bf 308}, 721 (1984)

\bibitem{Bouchaud97} E. Bouchaud, J. Phys. : Condens. Matter {\bf 9}, 4319 (1997) (and references therein).

\bibitem{Daguier95} P. Daguier {\em et al.}, Phys. Rev. Lett. {\bf 78}, 1062 (1997).

\bibitem{Nakano99} A. Nakano, R.K. Kalia and P. Vashishta, Phys. Rev. Lett. {\bf 75}, 3138 (1995);
P. Vashishta, R.K. Kalia, A. Nakano, Computing in Sci. and Eng., pp 56-65,  (september/october 1999); T. Campbell {\em et al.}, Phys. Rev. Lett. {\bf 82}, 4018 (1999).

\bibitem{vanBrutzel97} L. Van Brutzel, Ph.D. thesis (1999); L. Van Brutzel {\em et al.}, Mat. Res. Soc. Symp. Proc. {\bf 703}, V3.9.1 (2002); L. Rountree {\em et al.}, Annu. Rev. Mater. Res. {\bf 32}, 377 (2002).

\bibitem{Guilloteau96} E. Guilloteau, H. Charrue and F. Creuzet, Europhys. Lett {\bf 34}, 549 (1996);
S. Hénaux and F. Creuzet, J. Amer. Ceram. Soc. {\bf 83}, 415 (2000).

\bibitem{He95} M.Y. He, M.R. Turner  and A.G. Evans, Acta Materialia {\bf 43}, 3453 (1995).

\bibitem{Marliere02} C. Marli\`ere, F. Despetis and J. Phalippou, J. Non-Cryst. Solids (accepted).

\bibitem{Wiederhorn67} S.M. Wiederhorn, J. Amer. Ceram. Soc. {\bf 50}, 407 (1967);
S.M. Wiederhorn and L.H.  Bolz, J. Amer. Ceram. Soc. {\bf 53}, 543 (1970).

\bibitem{Marliere01} C. Marlière {\em et al.}, J. Non-Cryst.  Solids {\bf  85}, 148 (2001).

\bibitem{Kobayashi87} T. Kobayashi and D.A. Shockey, Metall. Transac. {\bf 18A}, 1941 (1987);
H. Miyamoto, M. Kikuchi and T. Kawazoe, Int. J. of Fract. {\bf 42}, 389 (1990)

\bibitem{Irwin57} G.R. Irwin, J. Appl. Mech. {\bf 24}, 361 (1957).

\bibitem{Lawn93} B.R. Lawn, {\em Fracture of Brittle Solids} (Cambridge University Press, Cambridge, 1993). [2nd edition]


\bibitem{Falk99} M.L. Falk, Phys. Rev. B {\bf 60}, 7062 (1999);M.L. Falk and J.S. Langer, MRS Bull. {\bf 25}, 40 (2000).

\bibitem{Bouchaud02} E. Bouchaud {\em et al.}, J. Mech. Phys. Solids {\bf 50}, 1703 (2002).

\bibitem{Benoit00} M. Benoit {\em et al.}, Eur. Phys. J. B {\bf 13}, (2000) 631; P. Jund, W. Kob and R. Jullien, Phys. Rev. B {\bf 64}, 134303  (2001).

\bibitem{Damien02} J.-C. Baret, D. Vandembroucq, S. Roux, Phys. Rev. Lett. {\bf 89}, 195506-1 (2002).


\end{thebibliography}
\end{document}